\documentstyle [prl,aps,floats]{revtex}

\begin{document}

\title{The Effect of the ($2+1$)-Dimensional Gravity in the Spiral Galaxy}
\author{W. F. Kao\footnote{E-mail address: {\tt wfgore@cc.nctu.edu.tw}} and
R.-T. Yu}
\address{Institute of Physics, Chiao Tung University, Hsin Chu, Taiwan}
\maketitle

\begin{abstract}
The slowly decreasing rotational velocity of the Milky Way suggests the existence of the dark matter.
One finds that the effect of the dark matter in a galaxy can be described by a $(2+1)$-dimensional fluid model.
The stability analysis for the corresponding fluid model is discussed in details in this paper.
It is found that the pressure term plays an important role in the stability of the spiral arms and counteracts the increasing mass of the observed galaxy as the radial distance increases. 
When the fluid region ends, the $ nearly $ constant rotational velocity of the distant stars require no additional mass if the spatial dimension in the remote region is close to two.
This indicates that the galaxies, we are interested, could be explained by a $(2+1)$-dimensional model. 
Possible implications are also discussed in this paper.
\end{abstract}
\vskip 0.5 cm


\section{Introduction}

It is known that our Milky Way looks like a thin-disk of shining stars with a diameter of roughly $8.5$ to $10\, kpc$ across \cite{galaxy}. 
The thickness of the Milky Way is roughly of the order of $100$ to $1000\, pc$. 
In fact, the diameter of the shining part for a typical spiral galaxy is roughly $10^6$ light year across.
This thin plane structure has a few problems needed to be resolved.
One of the issue is the possible existence of the dark matter \cite{dark,kolb} in order to account for its nearly asymptotically constant velocity along its spiral arms.

Indeed, the eccentric force of the farthermost stars in the spiral arms will require a huge interior mass to keep them on track in the family of the galaxy.
It is known that the ${1 / r^2}$ gravitational force will require a total mass $M(r)={v^2}r /{G} $ interior to the test star (with mass $m$) at radial distance $r$ in order to balance the eccentric force on the test star along its spiral arms.
Here $r$ is the radial coordinate in the spherical coordinates.

In fact, when the estimate was extended to distances beyond the point where the light from a galaxy ceases, it was found that $M(r)$ continues to increase.
On the other hand, the observed mass only provides roughly one percent or less \cite{kolb}.
This result is the conclusion based on a $1/r^2$ gravitational force.
Hence it is apparent that most of the mass source in our galaxy is dark in a $(3+1)$-dimensional spacetime.
This is also true for the entire universe, namely, the observed universe is closed to $\Omega=1$ Friedmann-Robertson-Walker space with most of its mass provider $dark$ throughout the universe \cite{kolb}. 
For a brief introduction of the dark matter please see, for example, chapter one of Ref. \cite{kolb}.
There are a number of candidates for the prescribed dark matter including the recently discovered massive neutrino detected at SuperK \cite{super}.

One notes that the effect of the $(3+1)$-dimensional dark matter in the spiral galaxy can be simulated by a $(2+1)$-dimensional fluid model.
This is because that the gravitational force for a point mass is proportional to ${1 / \rho}$ in a space with two spatial dimensions. Here $\rho$ denotes the radial coordinate in polar coordinates.
In other words, dark matter may not be needed if there is a $\ln \rho$ gravitational potential in effect in the scale of the order of the size of a typical galaxy.

Note that the $(2+1)$-dimensional fluid model will be introduced in section II. 
It will be shown that the pressure term present in the fluid model provides an expelling force that counteracts the increasing gravitational force due to the increasing mass along the spiral arms. 
In addition, isolated stars beyond the typical scale of the galaxy do not act as fluid any more.
Hence there does not exist a pressure term beyond that scale any more.
Fortunately, a $2$-dimensional $\ln \rho$ potential is enough to hold remote stars on track without requiring any additional interior mass.
Hence this model also explains the reason why there is no need for the dark matter beyond the shining region of the spiral galaxy.

In the following section, we will also show that spiral arm is a stable configuration for the $(2+1)$-dimensional fluid model without dark matter similar to the $(3+1)$-dimensional fluid model with additional dark matter for a wide range of pressure and rotational velocity.
Another way to interpret this result is that the $2$ spatial dimensional effect is induced by the $3$ spatial dimensional dark matter in a nontrivial way.
Some conclusions will be drawn at the end of this paper.

\section{Stability of a $(2+1)$-dimensional spiral galaxy}
Note that it was assumed that the mass density takes the form $\rho(x)=\sigma(\rho,
\varphi) \delta(z)$ for a $(3+1)$-dimensional fluid model in Ref. \cite{shu}.
We will assume a mass density of the form $\sigma(x)=\sigma(\rho,
\varphi)$ for a $(2+1)$-dimensional fluid model in this paper.
The Poisson equation reads
\begin{equation}
{1 \over \rho}\partial_\rho ( \rho \partial_\rho V) + {\partial_\varphi^2 \over
\rho^2}V = {4\pi \over 3} G \sigma.
\label{2ds}
\end{equation}
Furthermore, one has
\begin{eqnarray}
\partial_t \sigma + {\partial_\rho \over \rho} (\rho\sigma u)
+ {\partial_\varphi \over \rho^2} (\sigma j) &=& 0 , \label{2dt} \\
\sigma (\partial_t u + u \partial_\rho   u
+ {j \partial_\varphi \over \rho^2} u -{j^2 \over \rho^3}) &=& -\partial_\rho
\Pi - \sigma \partial_\rho V ,
\label{2dr} \\
\sigma (\partial_t j + u \partial_\rho   j
+ {j \partial_\varphi \over \rho^2} j ) &=& -\partial_\varphi \Pi
- \sigma \partial_\varphi V
\label{2dp}
\end{eqnarray}
from the conservation of energy and momentum in this $(2+1)$-dimensional fluid model.
Here the pressure term is parameterized as 
$\Pi \equiv \sigma v_t^2$ with $v_t$ the averaged thermal speed.
It is known that the pressure term has to be imposed as an external source \cite{shu}.
This term will play an important role for the stability of the $(2+1)$-dimensional fluid model for the spiral galaxy.

Assuming that 
$\Pi =\Pi(\sigma)$, one can expands the pressure term perturbatively following the expansion of 
$\sigma=\sigma_0 + \sigma_1$.
In addition, one will write $v_t(\sigma_0) = v_0$.
Here $|\sigma_1| \ll |\sigma_0|$ denotes a perturbation of density around the background mass density $\sigma_0$.
Furthermore, $u\equiv {d\rho / dt}$ and $j \equiv \rho^2 {d \varphi /
d t} = \rho^2 \omega(\rho)$ denote the radial velocity and angular momentum per unit mass respectively.

First assuming that there exists a stable background configuration 
$(\sigma, u, j, V) =(\sigma_0(\rho), 0, \rho^2\omega(\rho), V_0(\rho)\,)$, one can perform a small perturbation around this background.
One will also assume that the perturbed fields take the following form:
$(\sigma_*(\rho), u_*(\rho), j_*(\rho), V_*(\rho)
\, ) e^{i \chi(\rho,\varphi, t)}.$
Here $\chi(\rho,\varphi, t)= \Phi(\rho) +\omega_0t -m \varphi$ such that the existence of such kind of stable solution represents the existence of a stable spiral galaxy with $m$ arms.

Note however that $\omega_0$ is a complex angular velocity such with an imaginary part representing a decreasing or increasing factor of the perturbed field.
In addition, the WKBJ approximation $|\Phi(\rho)'| \gg |(\ln V_*)'|$ will be our standard approach \cite{shu}.
This simply means that the variational dependence of $\rho$ in  $V_*$ came mostly from the phase factor $\Phi(\rho)$.

One also assumes that \cite{shu}
$O({j_* / \rho}) \sim O(u_*)$, $ O({\sigma_0 /\rho}) \sim
O(\partial_\rho \sigma_0)$ and $O(\partial_\rho  \sigma_*) \sim O({\sigma_*
/ \rho})$ in addition to the WKBJ condition  $|\Phi(\rho)'| \gg |(\ln
V_*)'|$ in solving the equations of motion.
The consistency of the system requires that the following equation
\begin{equation}
(\omega_0 -m \omega)^2 =  k^2v_0^2 +
\Omega^2- {4\pi } G \sigma_0 /3
\end{equation}
remains valid from solving the system equations. 
Here $k \equiv {d\Phi / d\rho}$ in above expression.

Therefore the local stability of the spiral solution requires that
\begin{equation}
Q^2 \ge 1 -{Z^2 \over v_0^2} \label{Q2}
\end{equation}
from the consistency of the system equations.
Here $Q^2 \equiv {k^2 /
k_J^2}$, $Z^2 \equiv {\Omega^2 / k_J^2}$ with $k_J^2 \equiv
{4\pi G \sigma_0 /    (3 v_0^2)}$.
Therefore local stability is established whenever the constraint (\ref{Q2}) is obeyed. 
Hence the stability of this $(2+1)$-dimensional fluid model can be easily achieved similar to the $(3+1)$-dimensional fluid model.

\section{Conclusion}
We have shown that the dark matter in the Milky Way may not be required at all if a galaxy exhibits a $\ln \rho$ potential.
One hence goes further to show that the $(2+1)$-dimensional fluid model (without dark matter) does support stable spiral arms similar to the $(3+1)$-dimensional fluid model (with dark matter).
One also points out that the effective $(2+1)$-dimensional galaxy is probably nothing more than an effective theory induced by the effect of the dark matter in the $(3+1)$-dimensional fluid model.

Note however that most spiral galaxies are not exactly $(2+1)$-dimensional objects.
Hence the $(2+1)$-dimensional toy fluid model introduced here is only a simplified version.
One should consider a more realistic $(2^+ +1)$-dimensional fluid model in order to account for the formation of the physical galaxies.
The simplified model serves, however, to indicate that the dark matter can as well be replaced by other alternatives. 
Possible applications to the physics in the scale of the entire universe may also reveal more information indicated in this paper.
In addition, if the dark matter can not be found, one should pay more attention to the questions such as: what the large-scale physical dimensions are, and what dictates the physical process of the dimensional-reduction mechanism.

\begin{acknowledgments}

This work is supported in part by the National Science Council under the contract number NSC88-2112-M009-001.
\end{acknowledgments}

\end{document}